\documentclass[conference]{IEEEtran}
\usepackage{amsmath,amssymb,mathtools,amsfonts,mathrsfs,url,color,xcolor,colortbl,cite,graphicx,epstopdf,epsfig,tikz,pgfplots,lettrine,bigints,multirow}

\ifCLASSINFOpdf

\else

\fi

\newtheorem{lem}{\textbf{Lemma}}

\newenvironment{proof}[1][Proof]{\begin{trivlist}
\item[\hskip \labelsep {\bfseries #1}]}{\end{trivlist}}
\begin{document}

\title{Integrating UAVs into Existing Wireless Networks: A Stochastic Geometry Approach}

\author{Rabe Arshad$^1$, Lutz Lampe$^1$, Hesham ElSawy$^2$, and Md. Jahangir Hossain$^1$
       \\ \\
       $^1$University of British Columbia, Canada. $^2$King Fahd University of Petroleum and Minerals (KFUPM), KSA.\\
       Emails: rabearshad@ece.ubc.ca, lampe@ece.ubc.ca, Hesham.elsawy@kfupm.edu.sa, jahangir.hossain@ubc.ca
       \thanks{}
       }

\maketitle
\begin{abstract}
The integration of unmanned aerial vehicles (UAVs) into wireless networks has opened a new horizon to meet the capacity and coverage requirements foreseen in future wireless networks. Harnessing UAVs as flying base stations (BSs) has helped to achieve a cost-effective and on-the-go wireless network that may be used in several scenarios such as to support disaster response and in temporary hotspots. Despite the extensive research on the application side of UAVs, the integration of UAVs as BSs into existing wireless networks remains an open challenge. While UAV BSs can help to increase the area spectral efficiency, the added network heterogeneity and BS densification may diminish the capacity gains due to increased handover rates. In this work, we shed some light on this tradeoff by studying a three tier network architecture consisting of macro, small, and UAV BSs and analyze its coverage and rate performance. In particular, we consider joint control/data and split architectures and express the rate performance for stationary and mobile users, using an analytical approach based on stochastic geometry. While both joint and split architectures offer similar performance for stationary users, the numerical results suggest the splitting of control and data plane for mobile users.
\end{abstract}


\IEEEpeerreviewmaketitle

\section{Introduction}
A rapid proliferation of unmanned aerial vehicles (UAVs), commonly known as drones, is being observed worldwide. UAVs offer autonomous features, flexible reconfiguration, and variety of applications. Some electronic commerce companies like Amazon use UAVs to deliver packages to their customers while other technology giants like Facebook and X (formerly known as Google X) deploy UAVs to offer Internet services in areas with otherwise poor coverage. The increasing popularity of UAVs has motivated researchers to explore the opportunities and define the challenges for integrating UAVs into existing wireless networks (see \cite{survey_uav} for a survey and \cite{tutorial_uav} for a tutorial). This necessitates rigorous mathematical models for UAV aided networks to assess the actual foreseen performance.

UAV aided wireless networks are considered a key enabler to support diverse applications with orders-of-magnitude higher capacity requirements foreseen in future wireless networks. The continuous reduction in the cost of UAVs has made it cost-effective for the wireless operators to deploy UAV BSs in emergency situations and/or to complement existing networks. The wide range of operating altitude of UAVs suggests its usage for relaying transmission between two terrestrial BSs where direct line-of-sight (LOS) is not available \cite{uav_support,uav_relay}. Other applications may include deploying UAVs in hot-spot areas to meet capacity demand where the existing networks fail to support thousands of users at a time (e.g., concerts and sport events) \cite{uav_emergency}. Several studies are available in the literature, which analyzed the performance of UAV BSs. For instance, Mohamed \textit{et al.} \cite{optimal_uav} developed an algorithm for the optimal deployment of UAV BSs while balancing the tradeoff between coverage probability and transmit power. Boris \textit{et al.} \cite{UAV_PPP} modeled UAV BSs using Poisson point process (PPP) and studied their optimal heights to maximize the coverage probability. Zedenek \textit{et al.} \cite{uav_support2} analyzed the performance of UAVs both as relays and standalone BSs. They proposed that for a given scenario, several terrestrial BSs can be replaced by a single UAV BS while offering the similar rate performance. Sathyanarayanan \textit{et al.} \cite{uav_future} studied the design and implementation of a tethered Helikite (an optical fibre backhauled UAV BS) to provide Internet to the ground users. However, none of the aforementioned studies incorporated user mobility and the resulting handover (HO) rate effect into the analysis. The authors in \cite{uav_ho} studied a HO mechanism for aerial networks based on the adjustment of height and distance between UAV BSs. However, \cite{uav_ho} focusses on a single tier network and does not provide the interplay between HO rate and average throughput.
\begin{figure*}[!t]
\centering
\includegraphics[width=0.75\linewidth,keepaspectratio]{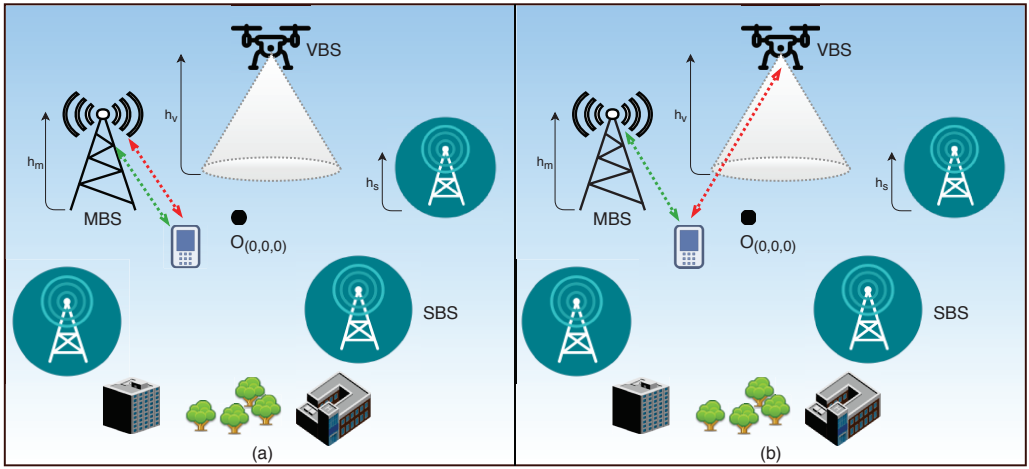} \vspace{-0.2cm}\caption{A three tier UAVs aided downlink network with (a) representing conventional mode and (b) representing C/D split mode. $h_k$, $k\mathrm{\in\{m,s,v\}}$ represent the heights of MBS, SBS, and VBS, respectively. The red and green lines represent control and data associations, respectively.}
\label{model}
\end{figure*}

Although the integration of UAV BSs into existing wireless networks, realized via heterogeneity and BS densification, tends to meet the desired spectral efficiency, the capacity gains are achieved at the expense of increased HO rates. Since the network densification shrinks the footprint of each BS, the mobile user experiences a relatively higher HO rate, which may diminish the foreseen capacity gains \cite{velocityaware,RabeAccess}. Also, the heterogeneity gives rise to the inter-tier HO rates that impose relatively higher service delays, and thus may further degrade the performance. Therefore, the integration of UAVs into existing networks necessitates the incorporation of HO effect into the network performance analysis. Furthermore, we need to minimize the HO effect in such highly dense heterogenous networks. In this paper, we consider a three tier network comprising macro BSs (MBSs), small BSs (SBSs), and UAV BSs (VBSs) and analyze the ground user experience in joint control/data (C/D) and  split architectures using tools from stochastic geometry. First, we consider a conventional three tier network with joint C/D plane and characterize coverage probability, HO rates, and average throughput. Second, since the HO rate in a highly dense heterogeneous network can be a performance limiting factor, we propose to split the control and data plane and assign the control management to the UAV BSs, which due to their high altitudes have the greater coverage. However, data is provided to the users based on the maximum received signal strength (RSS) rule. A recent study \cite{cu-split} showed that C/D splitting is beneficial for dense networks with mobile users, however, the work in \cite{cu-split} is limited to two tier networks. Moreover, \cite{cu-split} does not incorporate the BSs' heights into the mathematical analysis, which may significantly affect the network performance \cite{lower_height}. To the best of the authors' knowledge, no such study exists in the literature that presents a rigorous mathematical analysis incorporating the HO effect into the user rate analysis in a three tier 3-dimensional network. Furthermore, the C/D split architecture is exploited here to diminish the HO cost and enhance the user rate experience.

\section{System Model}
We consider a three tier downlink network comprised of MBSs, SBSs, and VBSs where the BSs belonging to each tier, $k\mathrm{\in\{m,s,v\}}$ are placed via a two-dimensional homogenous PPP $\Phi_{k}$ with intensity $\lambda_k$, antenna height $h_k$, and transmit with power $P_k$. The subscripts `m',`s',`v' denote the MBS, SBS, and VBS, respectively. The transmit power and height discrepancies among the tiers yield a weighted Voronoi tessellation~\cite{voronoi}. A Rayleigh fading environment with the universal frequency reuse and a power law path loss model with path loss exponent $\eta>2$ are considered. The users' locations are modeled via a homogeneous PPP with intensity $\lambda^{(u)}$. It is assumed that all users other than the test user are stationary. Without loss of generality, we conduct our analysis on a test ground mobile user following a random horizontal mobility model with velocity $v$. The mobile user changes its association as soon as it enters the Voronoi region of the target BS. A long user trajectory is considered so that the mobile user may pass through all association states. Similar assumptions are considered in ~\cite{HOben},\cite{HO_PCP}.

As mentioned earlier, we consider two network operating modes as shown in Fig.~\ref{model}. In the conventional mode, we assume that the control overhead consumes a fraction of $\mu_c$ of the overall bandwidth $W$. Also, the user receives the data and control signalling from the same BS. Here, we consider a universal frequency reuse and assume that all BSs transmit with the same frequency. This implies that the user receives interference from all BSs regardless of the tier. In the C/D split mode, UAV BSs are responsible to provide control signalling to the users regardless of their data associations. Thus we split the overall bandwidth\footnote{Although splitting overall bandwidth results in the spectral efficiency loss in the C/D split mode, the relatively lower handover cost and the elimination of control overhead from MBS/SBS mitigates this loss.} between UAV and other tiers to avoid interference between data and control signalling. For both operating modes, we calculate the average throughput experienced by a single user in an unloaded network as well as the average throughput per user in a fully loaded network.

\section{Average Throughput}
In this section, we calculate the average throughput experienced by a mobile user. Here, we are interested in the application throughput, which is obtained by subtracting signalling and control bits form the overall rate. The average throughput experienced by a typical mobile user in an unloaded three tier conventional and C/D split network can be expressed as
\begin{align}
AT^{(\mathrm{Con})}=\sum_{k}A_k T_k^{(\mathrm{Con})}(1-H_{c}^{(\mathrm{Con})}),
\label{AT1}
\end{align}
\vspace{-0.2cm}
\begin{align}
AT^{(\mathrm{Sp})}=\sum_{k}A_k T_k^{(\mathrm{\mathrm{Sp}})}(1-H_{c}^{(\mathrm{Sp})}),
\label{AT2}
\end{align}
where $A_k$ is the $k$th tier association probability, $T_k^{(.)}$ denotes the $k$th tier application rate, and $H_{c}^{(.)}$ represents the HO cost.
The average throughput per user in a fully loaded three tier conventional and C/D split network is given by
\vspace{-0.2cm}
\begin{align}
AT_{u}^{(\mathrm{Con})}=\sum_{k}\frac{A_k}{N_k} T_k^{(\mathrm{Con})}(1-H_{c}^{(\mathrm{Con})}),
\label{AT3}
\end{align}
\vspace{-0.2cm}
\begin{align}
AT_{u}^{(\mathrm{Sp})}=\sum_{k}\frac{A_k}{N_k} T_k^{(\mathrm{\mathrm{Sp}})}(1-H_{c}^{(\mathrm{Sp})}),
\label{AT4}
\end{align}
where $N_k$ denotes the mean number of users sharing the BSs resources with the typical user in the $k$th tier. The application rates $T_k^{(.)}$ are given by
\begin{align}
\label{application}
T_k^{(\mathrm{Con})}&=(1-\mu_c)W \mathbb{E}[\log_{2}(1+\mathrm{SINR}_k^{(\mathrm{Con})})],\\
T_j^{(\mathrm{Sp})}&=W_{c} \mathbb{E}[\log_{2}(1+\mathrm{SINR}_j^{(\mathrm{Sp})})], \quad j\in\mathrm{\{m,s\}},\\
T_\mathrm{v}^{(\mathrm{Sp})}&=(1-\mu_c^{\prime})W_{c}^{\prime} \mathbb{E}[\log_{2}(1+\mathrm{SINR}_\mathrm{v}^{(\mathrm{Sp})})],
\label{application1}
\end{align}
where $W=W_{c}+W_{c}^{\prime}$ and $\mu_c^{\prime}$ is the control overhead for the split architecture and is assumed to be greater than $\mu_c$ to meet control requirements for all users regardless of the serving tier.
The average number of users served by the $k$th tier ($N_k$) is computed as shown in \cite[Corollary 2]{users}:
\vspace{-0.2cm}
\begin{align}
N_k=\frac{1.28 \lambda^{(u)} A_k}{\lambda_k}+1.
\end{align}
The average spectral efficiency $\mathcal{R}=\mathbb{E}[\log_{2}(1+\mathrm{SINR}_k^{(.)})]$ in Eqs. \eqref{application} - \eqref{application1} can be computed using the coverage probability as follows \cite{cu-split}
\vspace{-0.2cm}
\begin{eqnarray} \label{rates}
\mathcal{R}&\stackrel{(a)}{=}& \int_{0}^{\infty}\mathbb{P}\left\{\log_{2}(1+\mathrm{SINR})>z\right\}dz,\\
&\stackrel{(b)}{=}&\int_{0}^{\infty}\frac{\mathbb{P}\left\{\mathrm{SINR}>T\right\}}{T+1}dT,
\label{SE}
\end{eqnarray}
where $(a)$ follows from the fact that $\log_{2}(1+\mathrm{SINR})$ is a strictly positive random variable and $(b)$ follows by the change of variables $T=e^{z}-1$.

In order to calculate the average throughputs given in Eqs. \eqref{AT1} - \eqref{AT4}, we first calculate the association probabilities followed by the distance distributions and coverage probabilities. Then we incorporate the HO effect by calculating HO rates for both network topologies.

\subsection{Association Probabilities}
In this section, we compute the association probabilities in a 3-dimensional three tier downlink network. Since the association probabilities depend on the relative BSs heights, we assume that $h_\mathrm{v} > h_\mathrm{m} >h_\mathrm{s}$. Let $Z_k$, $k\mathrm{\in\{m,s,v\}}$ be the Euclidean distance between the test user and the strongest BS belonging to the $k$th tier. The user associates with the SBS if $P_\mathrm{s}Z_\mathrm{s}^{-\eta}>P_\mathrm{m}Z_\mathrm{m}^{-\eta}$ and $P_\mathrm{s}Z_\mathrm{s}^{-\eta}>P_\mathrm{v}Z_\mathrm{v}^{-\eta}$. The association probabilities in a three tier network with BS heights constraints are given by the following lemma.
\begin{lem}
\label{assoc}
The association probabilities $A_k$, $k\mathrm{\in\{m,s,v\}}$ in a three tier network are shown on the next page in Eqs. \eqref{Am}-\eqref{Av}.
\begin{proof}
See Appendix A.
\end{proof}
\end{lem}
\begin{figure*}[!t]
\begin{multline}
A_\mathrm{m}=\underbrace{\frac{\lambda_\mathrm{m}}{\lambda_\mathrm{m}+\lambda_\mathrm{s} P_{\mathrm{sm}}}\Bigg(e^{-\pi\lambda_\mathrm{s}(P_{\mathrm{sm}}h_\mathrm{m}^{2}-h_\mathrm{s}^2)}-e^{-\pi P_\mathrm{mv}h_\mathrm{v}^{2}(\lambda_\mathrm{m}+\lambda_\mathrm{s} P_{\mathrm{sm}})+\pi(\lambda_\mathrm{m} h_\mathrm{m}^{2}+\lambda_\mathrm{s} h_\mathrm{s}^{2})}\Bigg)}_\text{$A_{\mathrm{m}_1}$} + \frac{\lambda_\mathrm{m}}{\lambda_\mathrm{m}+\lambda_\mathrm{s} P_{\mathrm{sm}}+\lambda_\mathrm{v} P_\mathrm{vm}}\cdot \\
\underbrace{e^{-\pi P_\mathrm{mv}h_\mathrm{v}^{2}(\lambda_\mathrm{m}+\lambda_\mathrm{s} P_{\mathrm{sm}}+\lambda_\mathrm{v} P_\mathrm{vm})+\pi(\lambda_\mathrm{m} h_\mathrm{m}^{2}+\lambda_\mathrm{s} h_\mathrm{s}^{2}+\lambda_\mathrm{v} h_\mathrm{v}^{2})}}_\text{$A_{\mathrm{m}_2}$}
\label{Am}
\end{multline}
\vspace{-0.3cm}
\begin{multline}
A_\mathrm{s}=\underbrace{1- e^{-\pi\lambda_\mathrm{s}(h_\mathrm{m}^{2}P_{\mathrm{sm}}-h_\mathrm{s}^{2})}}_\text{$A_{\mathrm{s}_1}$}+\underbrace{\frac{\lambda_\mathrm{s}}{\lambda_\mathrm{s}+\lambda_\mathrm{m} P_{\mathrm{ms}}}\Bigg(e^{-\pi P_{\mathrm{sm}}h_\mathrm{m}^{2}(\lambda_\mathrm{s}+\lambda_\mathrm{m} P_{\mathrm{ms}})+\pi(\lambda_\mathrm{s} h_\mathrm{s}^{2}+\lambda_\mathrm{m} h_\mathrm{m}^{2})}-e^{-\pi h_\mathrm{v}^{2}P_{\mathrm{sv}}(\lambda_\mathrm{s}+\lambda_\mathrm{m} P_{\mathrm{ms}})+\pi(\lambda_\mathrm{s} h_\mathrm{s}^{2}+\lambda_\mathrm{m} h_\mathrm{m}^{2})}\Bigg)}_\text{$A_{\mathrm{s}_2}$} + \\ \underbrace{\frac{\lambda_\mathrm{s}}{\lambda_\mathrm{s}+\lambda_\mathrm{m} P_{\mathrm{ms}}+\lambda_\mathrm{v} P_\mathrm{vs}}
e^{-\pi P_\mathrm{mv}h_\mathrm{v}^{2}(\lambda_\mathrm{m}+\lambda_\mathrm{s} P_{\mathrm{sm}}+\lambda_\mathrm{v} P_\mathrm{vm})+\pi(\lambda_\mathrm{m} h_\mathrm{m}^{2}+\lambda_\mathrm{s} h_\mathrm{s}^{2}+\lambda_\mathrm{v} h_\mathrm{v}^{2})}}_\text{$A_{\mathrm{s}_3}$}
\label{As}
\end{multline}
\vspace{-0.3cm}
\begin{align}
\hspace{-2.1cm}A_\mathrm{v}=\frac{\lambda_\mathrm{v}}{\lambda_\mathrm{v}+\lambda_\mathrm{m} P_\mathrm{mv}+\lambda_\mathrm{s} P_{\mathrm{sv}}}
e^{-\pi h_\mathrm{v}^{2}(\lambda_\mathrm{v}+\lambda_\mathrm{m} P_\mathrm{mv}+\lambda_\mathrm{s} P_{\mathrm{sv}})+\pi(\lambda_\mathrm{m} h_\mathrm{m}^{2}+\lambda_\mathrm{s} h_\mathrm{s}^{2}+\lambda_\mathrm{v} h_\mathrm{v}^{2})}, \quad \text{where } P_{kj}=\Big(\frac{P_k}{P_j}\Big)^{2/\eta},\forall k,j
\label{Av}
\end{align}
\hrulefill
\end{figure*}
The association probabilities given in Lemma \ref{assoc} are foundational to characterize the service distance distributions and coverage probabilities, which are given in the subsequent sections.

\subsection{Distance Distributions}
In this section, we calculate the service distance distributions that are required to determine the coverage probability and HO rates. Note that the distributions are calculated by assuming $h_\mathrm{v} > h_\mathrm{m} >h_\mathrm{s}$ and are given by the following lemma.
\begin{lem}
\label{PDF}
Let $X_k$, $k\mathrm{\in\{m,s,v\}}$ be the horizontal distance between the user and $k$th tier serving BS. Then the probability density functions (PDFs) of the distances between the user and the serving SBS, MBS, and VBS are given by
\begin{equation}
    f_{X_{\mathrm{m}}}(x) = \hspace{-0.1cm}\begin{cases}
        \frac{2\pi\lambda_\mathrm{m}}{A_{\mathrm{m}_{1}}} x e^{-\pi x^2(\lambda_\mathrm{m}+\lambda_\mathrm{s} P_{\mathrm{sm}})-\pi \lambda_\mathrm{s}(P_{\mathrm{sm}}h_\mathrm{m}^{2}-h_\mathrm{s}^{2}) },\\[-0.3em]\hspace{4.5cm} \text{ for } 0\leq x\leq L_{\mathrm{m}}\\
        \frac{2\pi\lambda_\mathrm{m}}{A_{\mathrm{m}_{2}}} x e^{-\pi x^2(\lambda_\mathrm{m}+\lambda_\mathrm{s}P_{\mathrm{sm}}+\lambda_\mathrm{v}P_{\mathrm{vm}})-\pi\lambda_\mathrm{s}( h_{\mathrm{m}}^{2}P_{\mathrm{sm}}- h_{\mathrm{s}}^{2})} \cdot\\[-0.3em] \hspace{1.8cm} e^{-\pi\lambda_\mathrm{v}( h_{\mathrm{m}}^{2}P_{\mathrm{vm}}- h_{\mathrm{v}}^{2})},\text{for }L_{\mathrm{m}} \leq x\leq \infty
        \end{cases}
        \label{mdist}
  \end{equation}
  \begin{equation}
    f_{X_{\mathrm{s}}}(x) = \hspace{-0.1cm}\begin{cases}
        \frac{2\pi\lambda_\mathrm{s}}{A_{\mathrm{s}_{1}}} x e^{-\pi\lambda_\mathrm{s}x^2},\hspace{0.5cm} \text{ for } 0\leq x\leq L_{\mathrm{s_1}}\\
        \frac{2\pi\lambda_\mathrm{s}}{A_{\mathrm{s}_{2}}} x e^{-\pi x^2(\lambda_\mathrm{s}+\lambda_\mathrm{m}P_{\mathrm{ms}})-\pi\lambda_\mathrm{m}( h_{\mathrm{s}}^{2}P_{\mathrm{ms}}- h_{\mathrm{m}}^{2})}, \\[-0.3em] \hspace{4.3cm} \text{for }L_{\mathrm{s_1}} \leq x\leq L_{\mathrm{s_2}}\\
        \frac{2\pi\lambda_\mathrm{s}}{A_{\mathrm{s}_{3}}} x e^{-\pi x^2(\lambda_\mathrm{s}+\lambda_\mathrm{m}P_{\mathrm{ms}}+\lambda_\mathrm{v}P_{\mathrm{vs}})-\pi\lambda_\mathrm{m}( h_{\mathrm{s}}^{2}P_{\mathrm{ms}}- h_{\mathrm{m}}^{2})} \cdot\\[-0.3em] \hspace{1.8cm} e^{-\pi\lambda_\mathrm{v}( h_{\mathrm{s}}^{2}P_{\mathrm{vs}}- h_{\mathrm{v}}^{2})},\text{for }L_{\mathrm{s_2}} \leq x\leq \infty
        \end{cases}
        \label{sdist}
        \end{equation}
        \begin{multline}
       f_{X_{\mathrm{v}}}(x)=\frac{2\pi\lambda_\mathrm{v}}{A_{\mathrm{v}}} x e^{-\pi x^2(\lambda_\mathrm{v}+\lambda_\mathrm{m}P_{\mathrm{mv}}+\lambda_\mathrm{s}P_{\mathrm{sv}})-\pi\lambda_\mathrm{m}( h_{\mathrm{v}}^{2}P_{\mathrm{mv}}- h_{\mathrm{m}}^{2})} \cdot\\[-0.3em] \hspace{1.8cm} e^{-\pi\lambda_\mathrm{s}( h_{\mathrm{v}}^{2}P_{\mathrm{sv}}- h_{\mathrm{s}}^{2})},\text{for }0 \leq x\leq \infty
        \end{multline}
        where $L_\mathrm{m}=\sqrt{h_\mathrm{v}^2P_\mathrm{mv}-h_\mathrm{m}^2}$, $L_{\mathrm{s_1}}=\sqrt{h_\mathrm{m}^2P_{\mathrm{sm}}-h_\mathrm{s}^2}$, and $L_{\mathrm{s_2}}=\sqrt{h_\mathrm{v}^2P_{\mathrm{sv}}-h_\mathrm{s}^2}$.
        \begin{proof}
        See Appendix B.
        \end{proof}
\end{lem}
\subsection{Coverage Probability}
In this section, we compute the coverage probabilities (i.e., $\mathbb{P}[\mathrm{SINR}>T]$) for the conventional and C/D split cases, which are then utilized to calculate the spectral efficiencies. The overall coverage probability is the weighted sum of coverage probabilities achieved through all types of associations and is given by
\vspace{-0.2cm}
\begin{align}
\mathcal{C}=A_{\mathrm{m}}C_\mathrm{m}+A_{\mathrm{s}}C_\mathrm{s}+A_{\mathrm{v}}C_\mathrm{v},
\label{CP}
\end{align}
where $C_{k}$, $k\mathrm{\in\{m,s,v\}}$ can be expressed as
\begin{align}
C_k=\mathbb{P}\left[\frac{P_k h Z_{k_1}^{-\eta}}{I_{agg}+\sigma^2}>T\right],
\label{cov}
\end{align}
where $h\sim \exp(1)$ represents the channel gain, $\sigma^2$ represents the noise power, $Z_{k_1}$ represents the distance between the user and the strongest/serving BS of $k$th tier, and $I_{agg}$ denotes the aggregate interference received from the other BSs/tiers except the serving BS. Note that \eqref{CP} holds for both conventional and C/D split architectures. However, the disparity comes in the aggregate interference $I_{agg}$. Let $I_k=\sum_{x\in \Phi_k\backslash b_{k_{1}}}^{}P_kh_xZ_{k_x}^{-\eta}$, $k\mathrm{\in\{m,s,v\}}$ be the interference observed from the $k$th tier, where $b_{k_1}$ represents the closest/serving $k$th tier BS belonging to $\Phi_k$ obtained via ordering the BSs w.r.t. the distances from the user. The interference in the conventional network comes from all tiers (i.e., $I_{agg}^{(con)}=I_{\mathrm{m}}+I_{\mathrm{s}}+I_{\mathrm{v}}$) due to the universal frequency recuse among all tiers. In the C/D split architecture, the MBS and SBS users do not receive interference from the UAV BSs as they use different operating frequencies. Therefore, the aggregate interference received by the MBS/SBS user in C/D split case is $I_{\mathrm{m}}+I_{\mathrm{s}}$. Also, the aggregate interference experienced by the VBS users in C/D split case will be $I_{\mathrm{v}}$. Similar assumptions are considered in \cite{cu-split} for a two tier network.

The exponential distribution of $h$ in Eq. \eqref{cov} leads to the conditional coverage probability as a function of Laplace transform (LT) of the aggregate interference as
\vspace{-0.1cm}
\begin{align}
C_k(X_{k_1})=\exp\left(-\frac{T \sigma^2 Z_{k_1}^{\eta}}{P_k}\right)\mathscr{L}_{I_{agg}}\left(\frac{T Z_{k_1}^{\eta}}{P_k}\right),
\label{cond}
\end{align}
where $Z_{k_.}=\sqrt{X_{k_.}^2+h_k^2}$. Since $I_{agg}$ is the summation of the independent interferences received from the individual tiers, the LT of the aggregate interference can be written as the product of the LTs of independent interferences. For instance, the LT of the aggregate interference in the conventional case can be written as
\vspace{-0.1cm}
\begin{align}
\mathscr{L}_{I_{agg}}(z)=\mathscr{L}_{I_{m}}(z)\mathscr{L}_{I_{s}}(z)\mathscr{L}_{I_{v}}(z).
\end{align}
The LTs of the aggregate interference in the conventional and split cases are given by the following lemma.
\begin{lem}
The LT of the aggregate interference in the conventional and split cases can be expressed in the terms of Gauss hypergeometric function~\cite{hypergeometric} as
\begin{multline}
\!\!\!\!\!\mathscr{L}_{I_{agg}}^{(\mathrm{Con})}\Big(\frac{T Z_{k_1}^{\eta}}{P_k}\Big)=\exp\!\Big\{\!-\!\frac{2\pi T Z_{k_1}^2}{\eta-2}\mathstrut_2 F_1\Big(1,1-\frac{2}{\eta},2-\frac{2}{\eta},-T\Big) \cdot \\
\Big(\lambda_k+\lambda_j
P_{jk}+\lambda_lP_{lk}\Big)\Big\}, \hspace{0.1cm}j,k,l \mathrm{\in\{m,s,v\}}, j\neq k\neq l,
\end{multline}
\begin{multline}
\!\!\!\!\mathscr{L}_{I_{agg}}^{(\mathrm{Sp})}\Big(\frac{T Z_{k_1}^{\eta}}{P_k}\Big)=\exp\Big\{-\frac{2\pi T Z_{k_1}^2}{\eta-2}\mathstrut_2 F_1\Big(1,1-\frac{2}{\eta},2-\frac{2}{\eta},-T\Big) \cdot \\
\Big(\lambda_k+\lambda_j
P_{jk}\Big)\Big\}, \hspace{0.1cm}j,k \mathrm{\in\{m,s\}}, j\neq k,
\label{sp1}
\end{multline}
\vspace{-0.2cm}
\begin{multline}
\!\!\!\!\!\mathscr{L}_{I_{agg}}^{(\mathrm{Sp})}\Big(\frac{T Z_{\mathrm{v}_1}^{\eta}}{P_k}\Big)=\exp\Big\{-\frac{2\pi\lambda_\mathrm{v} T Z_{\mathrm{v}_1}^2}{\eta-2}\cdot \\
\mathstrut_2 F_1\Big(1,1-\frac{2}{\eta},2-\frac{2}{\eta},-T\Big)\Big\},
\label{sp2}
\end{multline}
where Eq. \eqref{sp1} holds for the MBS/SBS users in the C/D split architecture where the interference is received from MBS/SBS only and Eq. \eqref{sp2} holds for the VBS users exhibiting interference received from the VBS tier only.
\begin{proof}
See Appendix C.
\end{proof}
\end{lem}
Finally, $C_k$ is obtained by substituting the LTs of the aggregate interference in Eq. \eqref{cond} and integrating it over the service distance distributions shown in Lemma \ref{PDF}. $C_k$ is then plugged into Eq. \eqref{SE} to obtain the spectral efficiency via numerical integration.
In what follows, we calculate the HO cost in the conventional and C/D split architectures, which along with the spectral efficiency will enable the computation of the average throughput.

\subsection{Handover Cost}
In this section, we compute the HO cost, a unitless metric, defined as the time required for performing a HO per unit time. Mathematically, it is given by
\begin{align}
H_{c}=d_{c}\cdot HO,
\end{align}
where $d_c$ represents the HO delay (i.e., time wasted in performing a HO) and $HO$ represents the HO rate or cell boundaries crossings per unit time. The overall HO costs in the conventional and C/D split architectures are given, respectively, by
\begin{align}
H_{c}^{(\mathrm{Con})}=d_{c}\sum_{i}\sum_{j}HO_{ij}, \quad i,j\mathrm{\in\{m,s,v\}},
\end{align}
\vspace{-0.2cm}
and
\begin{align}
H_{c}^{(\mathrm{Sp})}=d_{c}HO_{\mathrm{vv}}+ d_{c}^{\prime}\sum_{i}\sum_{j}HO_{ij\backslash i=j=\mathrm{v}}, \!\!\quad i,j\mathrm{\in\{m,s,v\}}.
\end{align}
Since the UAV BSs manage control signalling for all users in the C/D split case, the data HOs performed between MBS-MBS, SBS-SBS, and MBS-SBS do not exhibit long interruptions/delays. Moreover, the core network is not involved in the data HOs as the UAV BSs manage these HOs. Therefore, the data HO delays are assumed to be less than that of control HO i.e., $d_{c}^{\prime} < d_{c}$. The general HO rate expression for a multi-tier network is given by \cite{20a}
\begin{align}
HO_{ij}=\begin{cases}
\frac{2}{\pi}\mu(\mathcal{T}_{ij})v,\text{ for } i= j\\
\frac{1}{\pi}\mu(\mathcal{T}_{ij})v,\text{ for } i\neq j
\end{cases}
\label{Hkj}
\end{align}
\normalsize
where $\mu(\mathcal{T}_{ij})$ represents the length intensity of $ij$ cell boundaries (between tier $i$ and tier $j$), which is defined as the expected length of $ij$ cell boundaries in a unit square and $v$ represents the user velocity. Now we follow \cite{TVT} and exploit the association probabilities and distance distributions given in Lemma \ref{assoc} and Lemma \ref{PDF}, respectively, to compute the length intensities $\mu(\mathcal{T}_{ij})$ and finally the inter and intra tier HO rates $HO_{ij}$.

\begin{figure}[!t]
\centering
\includegraphics[width=0.85\linewidth,keepaspectratio]{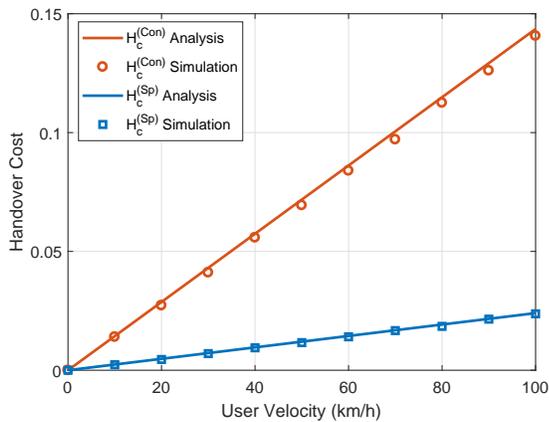} \vspace{-0.2cm}\caption{Handover cost versus user velocity}
\label{dho}
\end{figure}

\section{Numerical Results}
\begin{figure}[!t]
\centering
\includegraphics[width=0.85\linewidth,keepaspectratio]{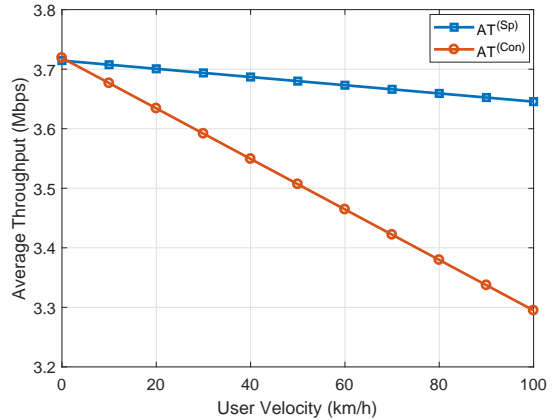} \vspace{-0.2cm}\caption{Average throughput versus user velocity}
\label{ATf1}
\end{figure}

\begin{figure}[!t]
\centering
\includegraphics[width=0.85\linewidth,keepaspectratio]{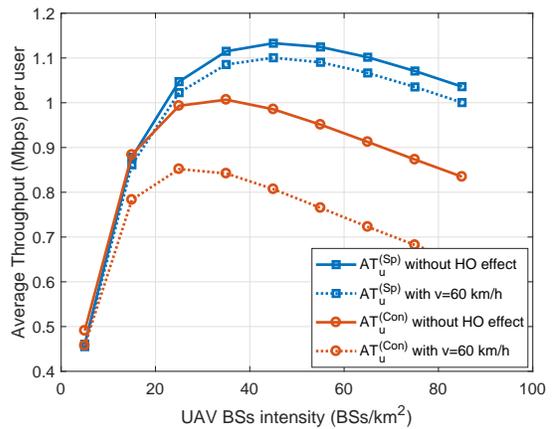} \vspace{-0.2cm}\caption{Average throughput per user versus UAV BS intensity with $\lambda_\mathrm{v}=2\lambda_\mathrm{m}=3\lambda_\mathrm{s}$}
\label{ATf2}
\end{figure}

In this section, we analyze the rate performance of mobile and stationary users in three tier conventional and C/D split network architectures with the network parameters shown in Table \ref{tab}. The HO cost for the conventional and split case is shown in Fig.~\ref{dho}. The trends show a considerable rise in the HO cost with the user velocity for the conventional case due to more frequent HOs when compared with the split case. Fig.~\ref{ATf1} shows the average throughput for a test user in an unloaded network versus user velocity. We observe a decrease in the average throughput due to a growing HO cost with the increase in the user velocity. While this is the case for both architectures, the results clearly show the advantage of splitting control and data planes for mobile users. The average throughput per user in a fully loaded network is shown in Fig.~\ref{ATf2}. The figure reveals the optimal UAV intensity that maximizes the average throughput per user. Moreover, the Fig.~\ref{ATf2} shows a turning point where the HO cost dominates the rate performance in a way that any further increase in the intensity leads to the degradation in the average throughput.
\begin{table}[!t]
\caption{\: Network parameters in accordance with \cite{3GPP_R14} where $k\mathrm{\in\{m,s,v\}}$}
\center
\resizebox{0.4\textwidth}{!}{
\begin{tabular}{|c c c c|}
\hline
\rowcolor{purple}
\multirow{-1}{*}{\textcolor{white}{\textbf{Parameter}}} & \multirow{-1}{*}{\textcolor{white}{\textbf{Value}} } &\multirow{-1}{*}{ \textcolor{white}{ \textbf{Parameter}}}  & \multirow{-1}{*}{\textcolor{white}{ \textbf{Value }}}  \\ \hline  \hline
& & & \\
 \multirow{-2}{*}{BS Power $P_k$:}             &  \multirow{-2}{*}{\{45, 24, 30\} dBm }    &  \multirow{-2}{*}{BS Height $h_k$:}  &   \multirow{-2}{*}{\{40, 20, 45\} m}     \\
 \cellcolor{purple!20!}  & \cellcolor{purple!20!}    &\cellcolor{purple!20!}    & \cellcolor{purple!20!}  \\
\multirow{-2}{*}{\cellcolor{purple!20!}BS Intensity $\lambda_k$:}  & \multirow{-2}{*}{\cellcolor{purple!20!}\{4, 15, 5\} BS/km$^2$} & \multirow{-2}{*}{\cellcolor{purple!20!}User Intensity $\lambda^{(u)}$:}  &   \multirow{-2}{*}{\cellcolor{purple!20!} 100 BS/km$^2$}   \\
& & & \\
 \multirow{-2}{*}{HO Delay $d_c$:}               & \multirow{-2}{*}{0.7 s}  &  \multirow{-2}{*}{HO Delay $d_{c}^{\prime}$:}       &  \multirow{-2}{*}{0.1 s}   \\
\cellcolor{purple!20!} & \cellcolor{purple!20!}  & \cellcolor{purple!20!}  &\cellcolor{purple!20!}   \\
 \multirow{-2}{*}{\cellcolor{purple!20!}Control Overhead $\mu_c$:}           &  \multirow{-2}{*}{\cellcolor{purple!20!} 0.3}     & \multirow{-2}{*}{\cellcolor{purple!20!}Control Overhead $\mu_{c}^\prime$:}     &  \multirow{-2}{*}{\cellcolor{purple!20!} 0.5} \\
 & & & \\
 \multirow{-2}{*}{SINR Threshold $T$:}               & \multirow{-2}{*}{0 dB}  &  \multirow{-2}{*}{Overall Bandwidth $W$:}       &  \multirow{-2}{*}{10 MHz} \\
 \cellcolor{purple!20!} & \cellcolor{purple!20!}  & \cellcolor{purple!20!}  &\cellcolor{purple!20!}   \\
 \multirow{-2}{*}{\cellcolor{purple!20!}Split Bandwidth $W_c$:}           &  \multirow{-2}{*}{\cellcolor{purple!20!} 7 MHz}     & \multirow{-2}{*}{\cellcolor{purple!20!}Path loss exponent $\eta$:}     &  \multirow{-2}{*}{\cellcolor{purple!20!} 4}

  \\ \hline
\end{tabular}
}
\label{tab}
\end{table}

\section{Conclusion}
In this paper, we consider a UAV aided three tier downlink network and study the rate performance for mobile and stationary users. In particular, we incorporate the effect of handover rates in the user rate and compare the rate performance in the joint C/D and split operating modes. The numerical results show comparable performance for the stationary users but advocate the usage of C/D split scheme for the mobile users. Moreover, a turning point is shown where the handover cost degrades the network performance despite increasing the BSs intensity.
\vspace{-0.21cm}
\appendices
\section{Proof of Lemma 1}
Let $R_\mathrm{m}=\sqrt{Z_{\mathrm{m}}^{2}-h_{\mathrm{m}}^{2}}$ be the horizontal distance between the MBS and the test user. Then the distribution of $Z_m$ can be calculated using the null probability of PPP and is given by
\vspace{-0.21cm}
\begin{align}
F_{Z_{\mathrm{m}}}(x)= 1-e^{-\pi\lambda_\mathrm{m} (x^2 - h_{\mathrm{m}}^2)}, \quad  h_{\mathrm{m}}<x<\infty,
\end{align}
Similarly, the distribution of $Z_s$ and $Z_v$ are can be expressed as
\vspace{-0.21cm}
 \begin{align}
 F_{Z_{\mathrm{s}}}(x)&=1-e^{-\pi\lambda_\mathrm{s} (x^2 - h_{\mathrm{s}}^2)},\quad h_{\mathrm{s}}<x<\infty,\\
 F_{Z_{\mathrm{v}}}(x)&=1-e^{-\pi\lambda_\mathrm{v} (x^2 - h_{\mathrm{v}}^2)},\quad h_{\mathrm{v}}<x<\infty.
\end{align}
\normalsize
  For the macro association probability $A_\mathrm{m}$, we first write
  \begin{align}
  A_\mathrm{m}&=\mathbb{P}[P_\mathrm{m} Z_{\mathrm{m}}^{-\eta}>P_\mathrm{s} Z_{\mathrm{s}}^{-\eta}, P_\mathrm{m} Z_{\mathrm{m}}^{-\eta}>P_\mathrm{v} Z_{\mathrm{v}}^{-\eta}]\notag \\
  &\stackrel{(c)}{=} \mathbb{E}_{Z_{\mathrm{m}}}\bigg\{\mathbb{P}[Z_{\mathrm{s}}>\sqrt{P_\mathrm{sm}} Z_{\mathrm{m}}]\cdot \mathbb{P}[Z_{\mathrm{v}}>\sqrt{P_\mathrm{vm}} Z_{\mathrm{m}}]\bigg\}
  \label{Ai}
   \end{align}

   \vspace{-0.1cm}
   where $(c)$ follows from the independence of $\Phi_\mathrm{s}$ and $\Phi_\mathrm{v}$.
   Then we solve \eqref{Ai} by assuming $h_\mathrm{m}<h_\mathrm{v}$ and exploiting the fact that $P[Z_{\mathrm{s}}>(\frac{P_\mathrm{s}}{P_\mathrm{m}})^{1/\eta} Z_{\mathrm{m}}]=e^{-\pi\lambda_\mathrm{s}[(\frac{P_\mathrm{s}}{P_\mathrm{m}})^{2/\eta}Z_{\mathrm{m}}^{2}-h_{\mathrm{s}}^2]}$ over the range $h_{\mathrm{m}}<Z_m<\infty$ and $P[Z_{\mathrm{v}}>(\frac{P_\mathrm{v}}{P_\mathrm{m}})^{1/\eta} Z_{\mathrm{m}}]=1$ over the range $h_{\mathrm{m}}\leq Z_m\leq h_{\mathrm{v}}\big(\frac{P_\mathrm{m}}{P_\mathrm{v}}\big)^{1/\eta}$ and $P[Z_{\mathrm{v}}>(\frac{P_\mathrm{v}}{P_\mathrm{m}})^{1/\eta} Z_{\mathrm{m}}]=e^{-\pi\lambda_\mathrm{v}[(\frac{P_\mathrm{v}}{P_\mathrm{m}})^{2/\eta}Z_{\mathrm{m}}^{2}-h_{\mathrm{v}}^2]}$ over the range $h_{\mathrm{v}}\big(\frac{P_\mathrm{m}}{P_\mathrm{v}}\big)^{1/\eta} \leq Z_m \leq \infty$. In case of SBS association, we first assume that $h_\mathrm{v}\big(\frac{P_\mathrm{s}}{P_\mathrm{v}}\big)^{1/\eta}>h_\mathrm{m}\big(\frac{P_\mathrm{s}}{P_\mathrm{m}}\big)^{1/\eta}$ and then exploit the fact that $P[Z_{\mathrm{m}}>(\frac{P_\mathrm{m}}{P_\mathrm{s}})^{1/\eta} Z_{\mathrm{s}}]=1$ over the range $h_{\mathrm{s}}\leq Z_s\leq h_\mathrm{m}(\frac{P_\mathrm{s}}{P_\mathrm{m}})^{1/\eta}$ and $P[Z_{\mathrm{m}}>(\frac{P_\mathrm{m}}{P_\mathrm{s}})^{1/\eta} Z_{\mathrm{s}}]=e^{-\pi\lambda_\mathrm{m}[(\frac{P_\mathrm{m}}{P_\mathrm{s}})^{2/\eta}Z_{\mathrm{s}}^{2}-h_{\mathrm{m}}^2]}$ over the range $h_\mathrm{m}(\frac{P_\mathrm{s}}{P_\mathrm{m}})^{1/\eta}\leq Z_s\leq \infty$. Also, $P[Z_{\mathrm{v}}>(\frac{P_\mathrm{v}}{P_\mathrm{s}})^{1/\eta} Z_{\mathrm{s}}]=1$ over the range $h_{\mathrm{s}}\leq Z_s \leq h_\mathrm{v}(\frac{P_\mathrm{s}}{P_\mathrm{v}})^{1/\eta}$ and $P[Z_{\mathrm{v}}>(\frac{P_\mathrm{v}}{P_\mathrm{s}})^{1/\eta} Z_{\mathrm{s}}]=e^{-\pi\lambda_\mathrm{v}[(\frac{P_\mathrm{v}}{P_\mathrm{s}})^{2/\eta}Z_{\mathrm{s}}^{2}-h_{\mathrm{v}}^2]}$ over the range $h_\mathrm{v}(\frac{P_\mathrm{s}}{P_\mathrm{v}})^{1/\eta}\leq Z_s\leq \infty$. Finally, the VBS association probability $A_\mathrm{v}$ is calculated by exploiting the fact that $P[Z_{\mathrm{m}}>(\frac{P_\mathrm{m}}{P_\mathrm{v}})^{1/\eta} Z_{\mathrm{v}}]=e^{-\pi\lambda_\mathrm{m}[(\frac{P_\mathrm{m}}{P_\mathrm{v}})^{2/\eta}Z_{\mathrm{v}}^{2}-h_{\mathrm{m}}^2]}$ and $P[Z_{\mathrm{s}}>(\frac{P_\mathrm{s}}{P_\mathrm{v}})^{1/\eta} Z_{\mathrm{v}}]=e^{-\pi\lambda_\mathrm{s}[(\frac{P_\mathrm{s}}{P_\mathrm{v}})^{2/\eta}Z_{\mathrm{v}}^{2}-h_{\mathrm{s}}^2]}$ over the range $h_\mathrm{v}\leq Z_v \leq \infty$.
   \vspace{-0.2cm}
\section{Proof of Lemma 2}
In order to calculate the PDF of the distance between the user and the serving BS, we first calculate the complementary cumulative distribution function (CCDF) given that the user associates with the $k$-th tier, $k\in\{\mathrm{m,s,v}\}$.
\vspace{-0.18cm}
\begin{eqnarray}
\mathbb{P}[X_{k}> x]=\mathbb{P}[R_{k}> x\big| n=k]=\frac{\mathbb{P}[R_{k}> x, n=k]}{\mathbb{P}[n=k]}
\label{A}
\end{eqnarray}
where $\mathbb{P}[n=k]=A_{k}$ represents the association probability of $k$th tier, which is given in Lemma~\ref{assoc}. Next, we calculate the joint distribution $\mathbb{P}[R_{k}> x, n=k]$ using the null probability of PPP. Then, we invoke the association probabilities conditioned on the heights in \eqref{A} and take the derivative w.r.t. $x$ to obtain the service distance distribution.
\section{Proof of Lemma 3}
The LT of the aggregate interference is calculated by following \cite[Lemma 2]{velocityaware}, assuming $\beta=1$, replacing $R_1$ by $Z_{k_1}$, and considering the appropriate interference for the conventional and C/D split architectures as mentioned above.
\bibliographystyle{IEEEtran}
\bibliography{IEEEabrv,mybibr}
\end{document}